\documentclass[prl,twocolumn,showpacs,preprintnumbers,amsmath,amssymb]{revtex4}

\usepackage{graphicx}
\usepackage{dcolumn}
\usepackage{bm}

\newcommand{\non}{\nonumber}
\newcommand{\rr}{{\bm r}}
\newcommand{\cdd}{c_{\rm dd}}

\newcommand{\ee}{{\bm e}}

\begin{document}

\preprint{APS/123-QED}


\title{Can Spinor Dipolar Effects be Observed in Bose-Einstein Condensates?}
\author{Yuki Kawaguchi$^1$}
\author{Hiroki Saito$^2$}
\author{Masahito Ueda$^{1,3}$}
\affiliation{$^1$Department of Physics, Tokyo Institute of Technology,
2-12-1 Ookayama, Meguro-ku, Tokyo 152-8551, Japan \\
$^2$Department of Applied Physics and Chemistry, The University of Electro-Communications,
1-5-1, Choufugaoka, Choufu-shi, Tokyo 182-8585, Japan \\
$^3$Macroscopic Quantum Control Project, ERATO, JST, Bunkyo-ku, Tokyo 113-8656, Japan
}
\date{\today}

\begin{abstract}
Weak dipolar effects in atomic Bose-Einstein condensates (BECs) have recently been predicted to develop spin textures.
However, observation of these effects requires magnetic field as low as $\sim 10~\mu$G for spin-1 alkali BECs, so that they are not washed out by the Zeeman effect.
We present a scheme to observe the magnetic dipole-dipole interaction in alkali BECs under a realistic magnetic field of $\sim 100$~mG.
Our scheme enables us to extract genuine dipolar effects and should apply also to $^{52}$Cr BECs.
\end{abstract}


\maketitle

One of the salient features of gaseous Bose-Einstein condensates (BECs) is the magnetic dipole-dipole interaction,
which is long-range, anisotropic, and exerts a tensor force.
The dipolar interaction is expected to yield rich phenomena when combined with spin degrees of freedom,
such as the Einstein--de Haas effect~\cite{Kawaguchi2006a,Santos2006} and
ground-state spin textures and mass currents~\cite{Yi2006,Kawaguchi2006b},
because dipole-induced spin textures can generate mass transport via spin-gauge symmetry~\cite{Ho1996}.
Unfortunately, however, the observation of these phenomena requires extremely low, if not unrealistic, magnetic field
typically below $10~\mu$G for spin-1 alkali BECs,
so that they are not masked by the Zeeman effect.
The only dipolar effect observed to date is the anisotropically distorted expansion of
spin-polarized $^{52}$Cr BEC~\cite{Stuhler2005}.

In this Letter, 
we show that a weak magnetic dipole-dipole interaction of atomic BECs induces
spin textures even in a magnetic field as high as 100~mG.
At such high magnetic field, the total magnetization of the system along the magnetic field is fixed by the Zeeman energy,
and therefore the predicted dipolar effects,
which rely on the spin relaxation, are unobservable~\cite{Kawaguchi2006a,Santos2006,Yi2006,Kawaguchi2006b}.
However, we show that the dipolar interaction can flip local spins
within the Hilbert subspace of a fixed total magnetization,
allowing spin textures to be generated.
This spin texture can be observed by spin-sensitive {\it in situ} imaging techniques~\cite{Higbie2005}
or by Stern-Gerlach separation.

We consider the dipole-dipole interaction between two magnetic dipole moments 
$\hat{\bm \mu}_1=g\mu_{\rm B}\hat{\bm F}_1^{\rm lab}$ and $\hat{\bm \mu}_2=g\mu_{\rm B}\hat{\bm F}_2^{\rm lab}$,
where $\hat{\bm F}_1^{\rm lab}$ and $\hat{\bm F}_2^{\rm lab}$ represent the hyperfine spin operators
in the laboratory frame,
$g$ is the Land\'{e} {\it g}-factor of the hyperfine spin, and $\mu_{\rm B}$ is the Bohr magneton.
The dipolar interaction between atoms located at $\rr_1$ and $\rr_2$ is given by
\begin{align}
\hat{v}_{\rm dd}(\rr_{12}) = \cdd
 \frac{\hat{\bm F}_1^{\rm lab}\cdot\hat{\bm F}_2^{\rm lab} - 3(\hat{\bm F}_1^{\rm lab}\cdot{\ee}_{12})(\hat{\bm F}_2^{\rm lab}\cdot{\ee}_{12})}{|\rr_{12}|^3},
\label{eq:dd_int}\end{align}
where $\rr_{12} = \rr_1-\rr_2$, $\ee_{12} = \rr_{12}/|\rr_{12}|$,
and $\cdd = \mu_0(g\mu_{\rm B})^2/(4\pi)$ with $\mu_0$ being the magnetic permeability of vacuum.
For spin-1 alkali atoms, we have $g=1/2$~\cite{Kawaguchi2006b}.

Suppose that a homogeneous external magnetic field $B$ is applied in the $z$ direction.
Since the linear Zeeman term $-\hat{\mu}_zB$ rotates the atomic spin around the $z$ axis at the Larmor frequency $\omega_{\rm L} = g\mu_{\rm B}B/\hbar$,
it is convenient to describe the system in the rotating frame of reference in spin space.
The spin operators $\hat{\bm F}_j$ $(j=1, 2)$ in the rotating frame are related to those in the laboratory frame by
$\hat{F}_{j\pm} \equiv \hat{F}_{jx}\pm i \hat{F}_{jy} = e^{\pm i\omega_{\rm L} t}\hat{F}_{j\pm}^{\rm lab}$ and $\hat{F}_{jz} = \hat{F}_{jz}^{\rm lab}$.
The dipolar interaction \eqref{eq:dd_int} can be rewritten in terms of $\hat{\bm F}$ as
\begin{align}
\hat{v}_{\rm dd}(\rr_{12})=&-\sqrt{\frac{6\pi}{5}}\frac{\cdd}{|\rr_{12}|^3} \non\\
\times &\bigg[ \frac{Y_{20}(\ee_{12})}{\sqrt{6}}\left(4\hat{F}_{1z}\hat{F}_{2z} - \hat{F}_{1+}\hat{F}_{2-} - \hat{F}_{1-}\hat{F}_{2+}\right) \non\\
&+Y_{2-1}(\ee_{12})\left( \hat{F}_{1+}\hat{F}_{2z}+\hat{F}_{1z}\hat{F}_{2+} \right)e^{-i\omega_{\rm L}t}\non\\
&-Y_{2 1}(\ee_{12})\left( \hat{F}_{1-}\hat{F}_{2z}+\hat{F}_{1z}\hat{F}_{2-} \right)e^{ i\omega_{\rm L}t}\non\\
&+Y_{2-2}(\ee_{12})       \hat{F}_{1+}\hat{F}_{2+}e^{-2i\omega_{\rm L}t}\non\\
&+Y_{2 2}(\ee_{12})       \hat{F}_{1-}\hat{F}_{2-}e^{ 2i\omega_{\rm L}t}\bigg],
\label{eq:vdd_rot}
\end{align}
where $Y_{2m}(\ee_{12})$ represents the $m$-th component of the rank-2 spherical harmonics.
When the Zeeman energy is much larger than the dipolar interaction energy,
i.e., $\hbar\omega_{\rm L} \gg \cdd |{\bm F}|^2/|{\bm r}_{12}|^3$,
the spin dynamics due to the dipolar interaction is much slower than the Larmor precession.
In a large magnetic field, 
we may therefore use an effective dipolar interaction which is time-averaged over the Larmor precession period $2\pi/\omega_{\rm L}$,
\begin{align}
\langle\hat{v}_{\rm dd}(\rr_{12})\rangle=& 
-\cdd\sqrt{\frac{\pi}{5}}\frac{Y_{20}(\ee_{12})}{|\rr_{12}|^3} \non \\
&\times\left(4\hat{F}_{1z}\hat{F}_{2z} - \hat{F}_{1+}\hat{F}_{2-} - \hat{F}_{1-}\hat{F}_{2+}\right).
\label{eq:vdd_ave}\end{align}
In contrast to Eq.~\eqref{eq:dd_int},
the time-averaged interaction~\eqref{eq:vdd_ave} separately conserves the projected total spin angular momentum and the projected relative orbital angular momentum.
However, the long-range and anisotropic nature of the dipolar interaction is still maintained in Eq.~\eqref{eq:vdd_ave}.
We will show below that $\langle \hat{v}_{\rm dd} \rangle$ generates spin texture
even in the case of weak magnetic dipole coupling of alkali atoms.

We study the spin dynamics in a system of $N$ spin-1 Bose-Einstein condensed atoms with mass $M$
confined in a spin-independent potential $U_{\rm trap}(\rr)$ in a magnetic field $B$.
In the rotating frame of reference in spin space,
the dynamics of the system is governed by the non-local Gross-Pitaevskii equation
\begin{subequations}
\begin{align}
i\hbar\frac{\partial\psi_{\pm1}}{\partial t} = &(H_0 + g_0n + q)\psi_{\pm1} + (g_1 f_{\mp} - g\mu_{\rm B} b_{\mp})\frac{\psi_0}{\sqrt{2}}& \non\\
 &\pm (g_1 f_z - g\mu_{\rm B} b_z)\psi_{\pm1},\\
i\hbar\frac{\partial\psi_0}{\partial t} = &(H_0 + g_0n)\psi_0 +(g_1f_+ - g\mu_{\rm B} b_+)\frac{\psi_1}{\sqrt{2}} \non\\
&+(g_1f_- - g\mu_{\rm B} b_-)\frac{\psi_{-1}}{\sqrt{2}},
\end{align}
\label{eq:GP}\end{subequations}
where $\psi_m(\rr,t)$ is the condensate wave function for the magnetic sublevel $m$
(which is related to that in the laboratory frame by $\psi_m^{\rm lab}(\rr,t)=\psi_m(\rr,t) \exp(im\omega_{\rm L} t)$),
$H_0=-\hbar^2\nabla^2/(2M) + U_{\rm trap}(\rr)$,
and $q=(\mu_{\rm B} B)^2/(4E_{\rm hf})$ is the quadratic Zeeman energy
with $E_{\rm hf}$ being the hyperfine energy splitting.
We define the number density $n=\sum_m|\psi_m|^2$,
longitudinal magnetization $f_z=|\psi_1|^2-|\psi_{-1}|^2$,
and transverse magnetization $f_+=f_-^*=f_x+if_y=\sqrt{2}(\psi_1^*\psi_0+\psi_0^*\psi_{-1})$.
The short-range interactions are characterized by
$g_0=4\pi\hbar^2(a_0+2a_2)/(3M)$ and $g_1=4\pi\hbar^2(a_2-a_0)/(3M)$
with $a_s$ $(s=0,2)$ being the {\it s}-wave scattering length for the scattering channel with total spin $s$.
We note that the linear Zeeman term is canceled in the rotating frame.

The time-averaged dipolar interaction $\langle \hat{v}_{\rm dd} \rangle$ induces
an effective dipolar field ${\bm b}(\rr)$ given by
\begin{align}
{\bm b}(\rr) &= C\int d\rr'\frac{Y_{20}(\ee)}{|\rr-\rr'|^3}\left[3f_z(\rr')\hat{z} - {\bm f}(\rr')\right],
\label{eq:b}
\end{align}
where $C=\sqrt{4\pi}\cdd/(\sqrt{5}g\mu_{\rm B})$, $\ee \equiv (\rr-\rr')/|\rr-\rr'|$, and $b_\pm = b_x \pm i b_y$.
It can easily be verified from Eqs.~\eqref{eq:GP} and \eqref{eq:b} that
the inhomogeneous Larmor precession around ${\bm b}(\rr)$ changes
the local longitudinal magnetization $f_z(\rr)$ as well as the local transverse magnetization $f_+(\rr)$,
while the total longitudinal magnetization $\int d\rr f_z(\rr)$ is conserved.

We perform numerical simulations of the spin dynamics
using the parameters of the Berkeley experiment~\cite{Higbie2005}.
In Ref.~\cite{Higbie2005}, a BEC of spin-1 $^{87}$Rb atoms was prepared in the $m=1$ state.
Then the spin was rotated to a transverse direction by applying a $\pi/2$ pulse of the rf field
and the ensuing Larmor precession of the transverse magnetization was probed
by spin-sensitive imaging.
The crucial observation to be discussed in detail below is that a significant dipolar effect can be observed
only if the spins are rotated through an angle other than $\pi/2$.

We consider a spin-1 $^{87}$Rb BEC of $N=4\times 10^6$ atoms 
in a trap with frequencies $(\omega_x, \omega_y, \omega_z)=2\pi(150, 400, 4)$~Hz
in a magnetic field of $B=54$ mG.
We first prepare a spin-polarized stationary state in $\psi_1$ by solving Eq.~\eqref{eq:GP}
with the imaginary-time propagation method while keeping $\psi_0=\psi_{-1}=0$,
and then rotate the spin state about the $y$ axis by a tilt angle $\theta$ at time $t=0$.
We simulate the ensuing spin dynamics by solving Eq.~\eqref{eq:GP} in three dimensions
with the Crank-Nicolson scheme.
The dipolar term is calculated by the convolution theorem
with a fast Fourier transform algorithm~\cite{Goral2002}.

The Thomas-Fermi (TF) radii of the BEC are $(R_x, R_y, R_z)=(6.7, 2.5, 252)~\mu$m
and the peak density is $n_0=5.6\times 10^{14}$~cm$^{-3}$,
which gives a dipole healing length of $\xi_{\rm dd}=\hbar/\sqrt{2M\cdd n_0}=5.7~\mu$m.
Since $R_x$ and $R_y$ are comparable to $\xi_{\rm dd}$,
the spin texture discussed below is almost one-dimensional in the $z$ direction~\cite{Kawaguchi2006b}.
We characterize the spin texture in terms of the integrated magnetization $\bar{\bm f}(z) = \iint dx dy {\bm f}(\rr)$.

Figure~1 shows the dynamics of texture formation for $\theta=40^{\tiny \rm o}$.
A schematic representation of spin texture is shown in Fig.~1(a).
Figures~1(b)--(e) show snapshots of the spin textures,
while Fig.~1(f) show the spin texture at $t=637$~ms calculated without the dipolar interaction.
Since no texture is developed in Fig.~1(f),
we conclude that the spin textures seen in Figs.~1(c)--(e) arise from the dipolar interaction.
The helix observed in Figs.~1(c)--(e) is symmetric with respect to $z=0$, i.e.,
it is right-handed for $z>0$ and left-handed for $z<0$.
A comparison of Fig.~1(f) with Fig.~1(b) shows that the quadratic Zeeman effect slightly rotates the spins.
Figure~1(g) shows the direction of the transverse magnetization $\phi=\arg(\bar{f}_+)$ as a function of $z$.
\begin{figure}
\includegraphics[width=\linewidth]{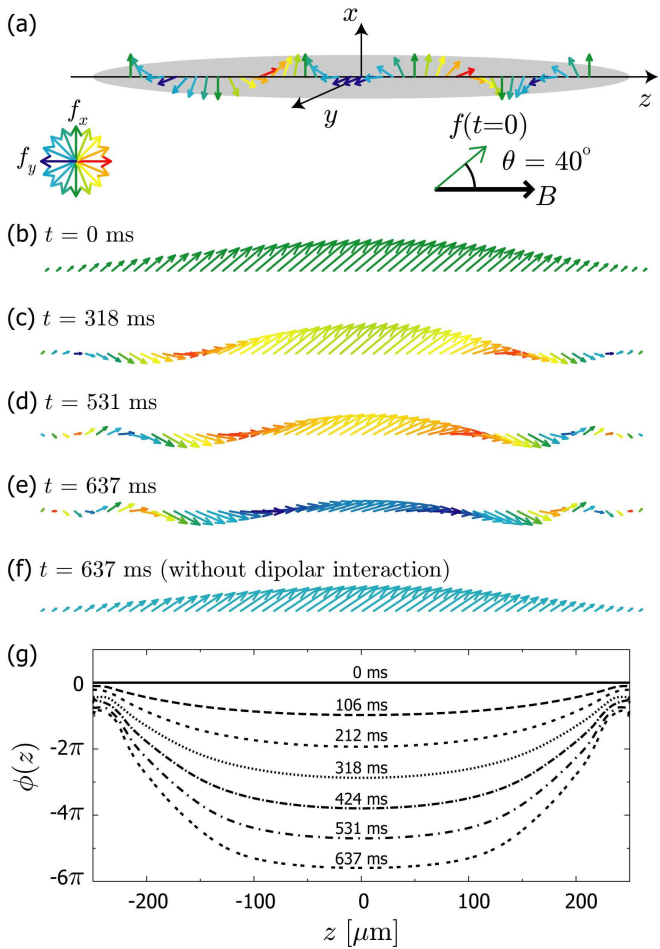}
\caption{(Color) (a) Schematic representation of spin texture.
The color of each arrow represents the azimuthal angle $\phi(z)=\arg[\bar{f}_+(z)]$ (see the color legend).
(b)--(e) Snapshots of spin texture seen from the $y$ direction
at (b) 0~ms, (c) 318~ms, (d) 531~ms, and (e) 637~ms after an rf pulse of $\theta=40^{\rm \tiny o}$.
(f) Snapshot of spin texture at 637 ms calculated without the dipolar interaction.
The arrows show the projection of $\bar{\bm f}(z)$ onto the $x$--$z$ plane.
The spin texture is displayed from $z=-214~\mu$m to $z=214~\mu$m.
(g) Time development of $\phi(z)$.
Note that $\phi(z)$ at $z\sim0$ becomes flat as time develops (see text).
}
\end{figure}

To evaluate the development of the spin texture,
we define the phase difference $\Delta\phi = \phi(0~\mu{\rm m})-\phi(155~\mu{\rm m})$ between $z=0~\mu$m and $z=155~\mu$m,
and plot the time development of $\Delta\phi$ for several tilt angles in Fig.~2(a).
Initially, $|\Delta\phi|$ increases linearly
and the spin texture develops more prominently for smaller $\theta$.
Since the second term in Eq.~\eqref{eq:b} is parallel to ${\bm f}(\rr)$ for uniformly polarized systems,
it does not initially contribute to the spin dynamics.
Therefore, the time dependence of $\phi$ at $t=0$ can be derived from Eqs.~\eqref{eq:GP} and \eqref{eq:b} as
\begin{align}
\frac{\partial \phi}{\partial t} &= -\frac{g\mu_{\rm B}}{\hbar}\bar{b}_0(z)\cos\theta,
\label{eq:phi-t}
\end{align}
where 
\begin{align}
\bar{b}_0(z) &= \frac{\iint dx dy n(\rr)\int d\rr' 3C\displaystyle\frac{Y_{20}(\ee)n(\rr')}{|\rr-\rr'|^3}}{\iint dxdy n(\rr)}
\label{eq:b0bar}
\end{align}
and we neglect the quadratic Zeeman effect.
We plot the $z$ dependence of $\bar{b}_0(z)$ in Fig.~2(b).
The time dependence of $\Delta\phi$ at $t\sim 0$ agrees well with Eq.~\eqref{eq:phi-t} as shown in the inset of Fig.~2(a),
where the black line corresponds to $\Delta\phi/\cos\theta = -(g\mu_{\rm B}/\hbar)\Delta \bar{b}_0t$
with $\Delta\bar{b}_0\equiv \bar{b}_0(0~\mu{\rm m}) -\bar{b}_0(155~\mu{\rm m})$, as indicated in Fig.~2(b).

It follows from Eq.~\eqref{eq:phi-t} that the dipole-induced spin texture does not appear when $\theta=90^{\rm \tiny o}$,
indicating that we can probe the inhomogeneity of the external field by setting $\theta=90^{\rm \tiny o}$.
Thus we can experimentally distinguish
the genuine dipolar effect from the one arising from residual magnetic-field gradient by changing $\theta$.

\begin{figure} 
\includegraphics[width=0.9\linewidth]{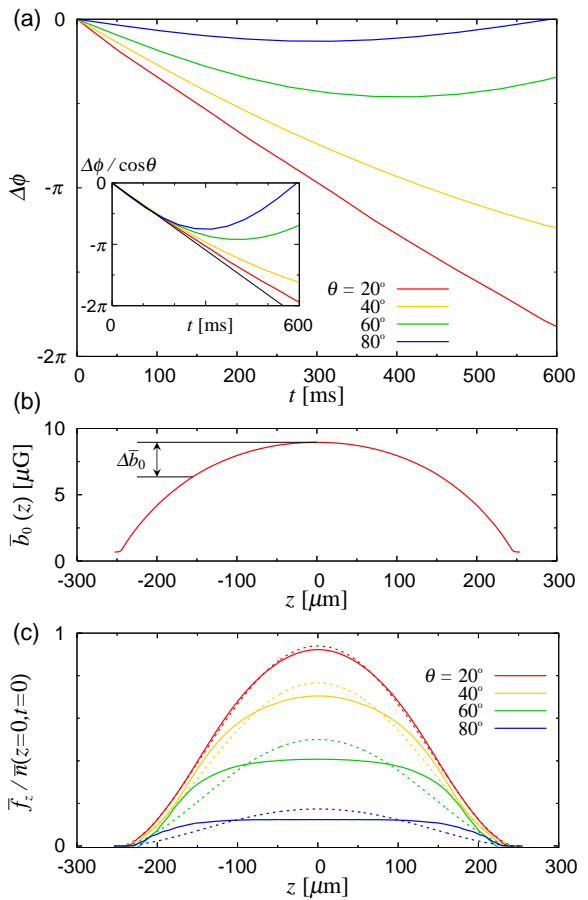}
\caption{(Color) 
(a) Time development of the phase difference $\Delta\phi$ between $z=0~\mu$m and $z=155~\mu$m.
The inset shows the same quantity scaled with $\cos\theta$.
The black line corresponds to $\Delta\phi/\cos\theta = -(g\mu_{\rm B}\Delta \bar{b}_0/\hbar)t$.
(b) $z$ dependence of the effective dipolar field at $t=0$ defined by Eq.~\eqref{eq:b0bar}.
(c) Profiles of the longitudinal magnetization $\bar{f}_z(z)$
at $t=0$~ms (dotted curves) and $t=531$~ms (solid curves)
scaled with the number density $\bar{n}(z=0,t=0)\equiv\iint dx dy n(x,y,z=0,t=0)$.
}
\end{figure}

As the spin texture further develops, the effective field cannot be approximated with $\bar{b}_0(z)$.
In fact, $\Delta\phi/\cos\theta$ deviates from the asymptotic line at $t\sim h/\epsilon_{\rm dd}=260$~ms,
where $\epsilon_{\rm dd}=2\pi \cdd n_0/3$ is the dipolar energy.
After $t\sim h/\epsilon_{\rm dd}$,
the spins tend to align in a high density region
and the helical structures appear only in low density regions,
so as to reduce the kinetic energy cost
[see the profile of $\phi(z)$ at $t=637$~ms in Fig.~1(g)].
Moreover, to reduce the energy cost of the helical structure,
the transverse magnetization decreases in the region where $|d\phi/dz|$ is large,
and instead the longitudinal magnetization increases in this region.
In Fig.~2(c), we plot the distribution of the longitudinal magnetization $\bar{f}_z(z)$ at $t=0$~ms (dotted curves) and $t=531$~ms (solid curves)
for several values of $\theta$.
We note that the distribution of $\bar{f}_z(z)$ is significantly deformed by the dipolar interaction, while $\int dz \bar{f}_z(z)$ is conserved.
After $t\sim600$~ms the spin texture develops into a complicated structure.

We have also simulated the texture formation in a pancake-shaped trap.
We consider a spin-1 $^{87}$Rb BEC of $N=5\times 10^6$ atoms in a magnetic field of $B=50$~mG
in a pancake-shaped trap with $(\omega_\perp,\omega_\parallel)=2\pi(4.3, 142)$~Hz,
where $\omega_\perp$ and $\omega_\parallel$, respectively, denote the trap frequencies perpendicular and parallel to the symmetry axis of the trap.
The TF radii of the BEC are $(R_\perp,R_\parallel)=(100, 3.0)~\mu$m
and the peak density is $n_0=1\times 10^{14}$~cm$^{-3}$, giving $R_\parallel < \xi_{\rm dd}=13~\mu{\rm m} < R_\perp$.
Figure~3 shows snapshots of the spin texture at $t=749$~ms
after rotation of the atomic spin about the $y$ axis by $\theta=60^{\rm \tiny o}$.
The symmetry axis of the trap is taken to be parallel [Fig.~3(a)] and perpendicular [Fig.~3(b)] to the magnetic field ($B\parallel z$).
Because of the anisotropy of the dipolar interaction,
the obtained spin texture strongly
depends on the direction of the trap anisotropy.
When the magnetic field is parallel to the symmetry axis, the spin texture is concentric in the $x$--$y$ plane [Fig.~3(a)].
On the other hand, when the magnetic field is applied perpendicular to the symmetry axis,
the spin texture is highly anisotropic [Fig.~3(b)] because
the effective dipolar field is not symmetric with respect to the symmetry axis of the trap.
\begin{figure}
\includegraphics[width=\linewidth]{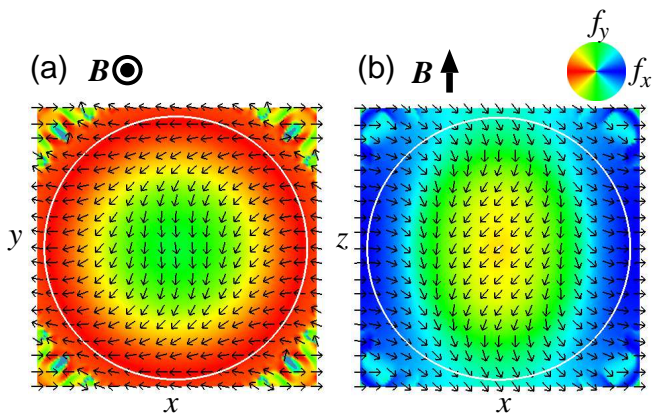}
\caption{(Color) Snapshots of spin texture in a pancake-shaped trap at $t=749$~ms after an rf pulse of $\theta=60^{\rm \tiny o}$.
The magnetic field is (a) parallel and (b) perpendicular to the symmetry axis.
The color shows the direction of the transverse magnetization $\phi=\arg(\bar{f}_+)$,
where the overbar denotes an integrated value along the symmetry axis.
The white circles show the TF radius and black arrows 
show transverse magnetization $(\bar{f}_x, \bar{f}_y)/\bar{n}$,
where the black arrows in (b) represent $(\bar{f}_x, \bar{f}_y)/\bar{n}$ on the $x$--$z$ plane of the coordinate space
with $\rightarrow$ and $\uparrow$ indicating $(\bar{f}_x, \bar{f}_y)/\bar{n}=(1,0)$ and $(0,1)$, respectively.
}
\end{figure}

For $B \sim 50$~mG, as used in the above calculations,
the quadratic Zeeman energy ($\sim 0.01$~nK) is much less than the dipolar interaction energy ($\epsilon_{\rm dd}\sim 0.2$~nK),
and hardly affects the texture formation.
Using the parameters from the Berkeley experiment~\cite{Higbie2005},
we found that the results are qualitatively insensitive to the magnetic field for $B\lesssim200$~mG.

The texture-formation dynamics can be observed also in other atomic species.
We have obtained qualitatively the same results for $^{23}$Na and $^{52}$Cr atoms 
as those for $^{87}$Rb:
transverse spin texture first develops due to the inhomogeneity of $\bar{b}_0$,
followed by a change of the local longitudinal magnetization.
The time scale of the texture formation depends on the strength of the dipole-dipole interaction;
thus, for the case of $^{52}$Cr atoms with a magnetic dipole moment of $6\mu_{\rm B}$, spin texture develops 144 times faster than for spin-1 alkali atoms.

In conclusion, we have shown that
the dipole-dipole interaction between atomic dipole moments can induce observable effects
on spin dynamics under an experimentally accessible magnetic field ($\sim 100$~mG),
even though the short-range interaction and Zeeman energy dominates the dipolar interaction.
Our proposal can be verified by taking a tilt angle $\theta$ of the spin other than $90^{\rm \tiny o}$ in
the Berkeley experiment~\cite{Higbie2005}.

\begin{acknowledgments}
This work was supported by Grants in Aid for Scientific Research (Grants
No.\ 17071005 and No.\ 17740263) and by 21st Century COE
programs on ``Nanometer-Scale Quantum Physics'' and ``Coherent Optical Science''
from the Ministry of Education, Culture, Sports, Science and Technology of Japan.
YK acknowledges support by the Japan Society for 
the Promotion of Science (Project No.\ 185451).
MU acknowledges support by a CREST program of the JST.
\end{acknowledgments}



\end{document}